\documentclass[12pt]{article}
\usepackage[a4paper,width=6.8in,height=8.8in]{geometry}
\usepackage{amsmath,amssymb}

\usepackage{amsmath,amssymb,mathtools,amsthm,
	textcomp,mathscinet}
\usepackage{amsfonts,graphicx,enumerate,subcaption,placeins}
\usepackage{longtable}
\usepackage{setspace}
\usepackage[utf8]{inputenc}
\usepackage[english]{babel}
\usepackage{booktabs,multirow}
\usepackage{blkarray}
\usepackage{lscape}
\usepackage[T1]{fontenc}
\usepackage{authblk,color}
\usepackage[english]{babel}
\usepackage{amsfonts, amstext, amsthm, booktabs, dcolumn}
\usepackage{graphicx}
\usepackage{float}
\usepackage{caption}
\usepackage{centernot}
\usepackage{times}
\usepackage[nottoc]{tocbibind}
\usepackage[mathscr]{eucal}

\usepackage[colorlinks=true,linkcolor=blue, allcolors=blue]{hyperref}%

\definecolor{gr}{rgb}{0.7, 0.0, 0.15}

\numberwithin{equation}{section}
\begin{document}
%\title{Statistical Modelling of High-Frequency Cryptocurrency data}
\title{Prediction of Cryptocurrency Prices through a Path Dependent Monte Carlo Simulation}
\author[1]{{\Large Ayush Singh}}
\author[2]{{\Large Anshu K. Jha}}
\author[*]{{\Large Amit N. Kumar}}
\affil[ ]{Department of Mathematical Sciences, Indian Institute of Technology (BHU),}
\affil[ ]{Varanasi (Uttar Pradesh) - 221 005, India.}
\affil[1]{Email: ayush.singh.2k26@gmail.com}
\affil[2]{Email: anshujha271@gmail.com}
\affil[*]{Email: amit.mat@iitbhu.ac.in}

\date{}
\maketitle

\begin{abstract}
\noindent
In this paper, our focus lies on the Merton's jump diffusion model, employing jump processes characterized by the compound Poisson process. Our primary objective is to forecast the drift and volatility of the model using a variety of methodologies. We adopt an approach that involves implementing different drift, volatility, and jump terms within the model through various machine learning techniques, traditional methods, and statistical methods on price-volume data. Additionally, we introduce a path-dependent Monte Carlo simulation to model cryptocurrency prices, taking into account the volatility and unexpected jumps in prices.
\end{abstract}

\noindent

\section{Introduction}
The cryptocurrency market behaves quite differently from the traditional stock market. It operates continuously around the clock and shows signs of extreme volatility. Notably, it lacks market makers and experiences low liquidity, factors that contribute to its increased volatility and susceptibility to sudden price jumps. For instance, the total market capitalization of Bitcoin was \$1.2 billion in May 2013, which grew to \$100 billion mark in October 2017 and briefly reached its highest market capitalization of \$1.28 trillion mark in November 2021. This unique behavior has garnered significant interest in recent times, making cryptocurrency markets an intriguing area of study for researchers and investors. Several researchers have studied cryptocurrency price forecasting, employing various methodologies and data sources. The investigation into modeling cryptocurrency prices has garnered special attention due to the sudden surge of interest in cryptocurrencies. Multiple statistical models, including Auto-Regressive Integrated Moving Average (ARIMA) and Generalized Auto Regressive Conditional Heteroskedasticity (GARCH), have been investigated to capture price trends in cryptocurrency markets. Additionally, researchers have explored the role of social media sentiment and trading volume in forecasting cryptocurrency prices, underscoring the impact of market sentiment and user behavior on price fluctuations. For more details, see Chokor and Alfieri \cite{regulationcrypto}, Katsiampa \cite{GarchBtc}, Lamon {\em et al.} \cite{socialbtc2}, Phillips and Gorse \cite{socialbtc}, Poongodi {\em et al.} \cite{arima_btc}, and reference therein.

\noindent
Several Machine Learning (ML) algorithms have been explored in the context of price prediction across various markets such as the stock market (Basak {\em et al.} \cite{stockml1}), gold market (Manjula and Karthikeyan \cite{goldML1}, and Livieris {\em et al.} \cite{goldDL1}), crude oil market (Abdullah and Zeng \cite{OilML1}), and futures and options (Culkin and Das \cite{OptionsML1}), among others. Numerous studies have aimed to understand Bitcoin's price trend and forecast its future value using ML algorithms. Other studies have explored the utilization of ML models (Chen \cite{BITCOINML2} and McNally {\em et al.} \cite{BITCOINML1}) and Deep Learning Models (Rizwan {\em et al.} \cite{BITCOINDL1} and Shin {\em et al.} \cite{BITCOINDL2}), both of which have surpassed traditional statistical models such as ARIMA and GARCH. In particular, Chen \cite{BITCOINML2} demonstrated in his research that the prediction accuracy and error of random forest models outperform those of the LSTM model. Furthermore, the correlation between the price of Bitcoin and other assets has been studied by several researchers such as Aggarwal {\em et al.} \cite{goldbtc} (gold market), Selmi {\em et al.} \cite{oilbtc} (crude oil market), among others. For instance, Sifat {\em et al.} \cite{leadlagbtc} demonstrated in their research a correlation between the prices of Bitcoin and Ethereum, albeit without clear discernible patterns in their modeling. Moreover, Tarnopolski \cite{montebtc} has studied empirical simulations of Bitcoin's price using methods including geometric Brownian motion and Monte Carlo simulations. This paper aims to study and provide a new way to model cryptocurrency prices, bearing in mind the volatility and unexpected jumps in prices.

\noindent
The Merton jump diffusion (MJD) model, introduced by Merton \cite{MERTON}, is a mathematical framework used to describe the movement of asset prices over time, particularly in financial markets. It extends the basic Black-Scholes model by incorporating jumps, which are sudden and significant changes in asset prices that occur infrequently but have a notable impact on the overall price dynamics. Applying the Lee and Mykland \cite{LM} jump tests to the $k$-line spot data from Binance of bitcoin on a 1-minute frequency across the entire period of 2021 and 2022, we noticed an average of 3.5 jumps per day, over the period. The MJD model comprises three primary components: drift, volatility, and the jump process. Given the extreme volatility and unexpected jumps observed in cryptocurrency markets, which can have substantial impacts on stock prices, our objective is to employ Merton's jump diffusion process for modeling purposes. In this paper, we use Merton's jump diffusion model, modeling the drift, diffusion, and jump terms via various methodologies such as price modeling using metrics from OHLCV data, order book data, forecasting using various ML techniques. We integrate various ML techniques, traditional methods, and price-volume data to model the prices of different cryptocurrencies, including Bitcoin and Ethereum, among others, using Merton's jump diffusion model as the foundational framework incorporating various changes in the jump, drift, and diffusion terms. We compare the results over a period spanning three years from January 2020 to January 2023, utilizing data sourced from the popular cryptocurrency exchange, Binance. Finally, we introduce a path-dependent Monte Carlo simulation in Python aimed at simulating the future distribution of spot cryptocurrency prices. This simulation leverages the jump diffusion process with drift, offering a novel approach to predict future price movements in cryptocurrency markets.

\noindent
Monte Carlo simulation, introduced by Metropolis and Ulam \cite{montecarlo}, is a versatile tool employed in various diverse fields such as finance, engineering, and physics for modeling complex systems and making predictions under uncertainty. This method uses random sampling to approximate system behavior, generating numerous potential outcomes by iteratively sampling from input variable probability distributions. In finance, Monte Carlo simulations are extensively utilized to model stock prices, interest rates, and other financial instruments, proving particularly valuable in options pricing, risk management, and portfolio optimization due to their ability to incorporate uncertainty and randomness. In this paper, we aim to divide our approach into categories of regression and classification prediction for our final results post-modeling with Monte Carlo Simulations. The Monte Carlo simulation provides an estimate of the Bitcoin price, which we utilize as the basis for our regression and classification analysis. We use the estimated price data to generate signals for long or short positions, which form the basis of our classification problem. To evaluate our results, we make use of metrics such as Mean Absolute Error (MAE), Root Mean Squared Error (RMSE), and Mean Absolute Percentage Error (MAPE) for regression analysis. For classification-based results, we use metrics such as Binary Accuracy, F1-Score, Precision, Recall, and Sensitivity. 

\noindent
This paper is organized as follows. In Section \ref{sec: Methodology}, we present various methodologies for predicting the drift and volatility of MJD model. Additionally, we discuss metrics that are crucial for deriving our primary findings. In Section \ref{12:sec3}, we address data availability and preprocessing, including statistical features of data, distribution of price change factors and volatility terms. In Section \ref{12:sec4}, we introduce an algorithm for simulating data and presents Monte Carlo simulations for BTCUSDT pair price paths within MJD model. Finally, in Section \ref{12:sec6}, we delve into our findings and discuss relevant remarks.

%Chen and Guestrin \cite{xgboost} propose an algorithm for space data and weighted quantile sketch for approximate tree learning, and

\section{Methodology} \label{sec: Methodology}
The MJD model serves as the foundation of our methodology. It is defined by the following stochastic differential equation (SDE): 
\begin{align}
\frac{dS(t)}{S(t)} = \mu dt + \sigma dW(t) + dJ(t), \label{eq:3.1}
\end{align}
where $S(t)$ denotes the price of the asset at time $t$, $\mu$ is the drift term,  $\sigma$ is the volatility of the asset or the diffusion term, $W(t)$ is a Wiener process, and $J(t)$ represents the jump process. We now divide the MJD model into three distinct tasks: drift prediction, volatility prediction, and jump term analysis. Each task is addressed individually, employing various techniques. Throughout this paper, the jump parameter $J(t)$ is characterized by a compound Poisson process, defined by the stochastic process:
\begin{align}
J(t) = \sum_{i=1}^{N(t)} Y_i ,\label{eq:2.2}
\end{align}
where $N(t)$ is a Poisson process with rate $\lambda$, $Y_i\sim N(m, s^2)$, the normal distribution with mean $m$ and variance $s^2$, and $Y_i$'s are \textit{iid} (independent and identically distributed). It can be easily verified that
\begin{align}
    J(t) \sim N(m \lambda t, \lambda t s^2).   \label{eq:2.3}  
\end{align}
The numerical solution of \eqref{eq:3.1} is given by 
\begin{align}
S(t) = S(0) e^{\left( \mu - \frac{1}{2} \sigma^2 \right) t + \sigma W(t) + J(t)}.\label{eq:3.4}
\end{align}
We simulate the above solution using the Monte Carlo simulation technique. Therefore, the solution now has two unknowns $\mu$ and $\sigma$. We forecast and estimate $\mu$ and $\sigma$ using various ML and empirical methodologies and compare the results across all possible combinations of the models. For more details, see Merton \cite{MERTON}, Kou \cite{jumpdiffpoiss}, and reference therein. \\
Next, we outline the techniques that can be employed to model the drift coefficient.

\subsection{Drift Modeling}
In the MJD model, the drift term $\mu$ encompasses the risk-free interest rate, the mean jump size, and the intensity of the jumps.  However, our focus lies solely on forecasting $\mu$ that represents the mean of the drift term. To forecast the drift rate, we employ a range of techniques, including ML models (linear regression and polynomial regression), eXtreme Gradient Boosting (XGBoost) regression, and Light Gradient Boosting Machine (LGBM) regression, as well as methods such as percent change (PC) in price over previous periods and the Drift of Percent Change (DoPC) of price over previous periods (rolling mean with fixed intervals of percent change). For comparative analysis, we consider the DoPC as our base model against which we evaluate the performance of our ML models across various metrics. The PC in price and DoPC can be defined as follows:
\begin{align}
\delta_i =  \frac{(P_i - P_{i-1})}{P_{i-1}} \quad \text{and}\quad  \rho_i = \frac{1}{n}\sum_{j=i-n}^{i} \delta_j ,\label{eq:3.5}
\end{align}
where  $P_i$ denotes the price of Bitcoin at the start of the $i$-th hour, $\delta_i$ represents PC in price between $(i-1)$-th and $i$-th hour, $n$ denotes the number of periods in our rolling drift calculation, and $\rho_i$ represent the drift of percent change. We define the delta of drift as 
\begin{align}
    \Delta_i := \delta_i - \delta_{i-1}.\label{eq:3.7}
\end{align}

% Where $\mu$ is the drift, $n$ is the number of observations (Hour returns), and $r_i$ represents the $i$-th Hour return.

% ML Models
\noindent
ML models have garnered a lot of interest in various forecasting problems. Our objective is to employ a range of ML models to substitute the drift term with the forecasted drift term. Specifically, we leverage linear regression, polynomial regression, XGBoost regression, and LGBM regression techniques. We develop a model of each type to predict the drift value for the upcoming month at the onset of each month, utilizing training data from the preceding month. For more details, we refer the reader to Chen {\em et al.} \cite{xgboost_btc}, Sun {\em et al.} \cite{lgbmbtc}, Uras {\em et al.} \cite{lr_gbm_btc}, and reference therein.

\subsubsection{Linear Regression}\label{subsec:3.1.1}
Linear regression is a statistical method employed to establish the connection between a dependent variable and one or more independent variables. Our objective is to model the dependent variable with the independent variables under the assumption of a linear relationship. Mathematically, this relationship is expressed as follows:
\begin{align*}
    \boldsymbol{y} = \beta_0 + \beta_1 \boldsymbol{x_1}+\beta_2 \boldsymbol{x_2} + \cdots + \beta_n \boldsymbol{x_n}.
\end{align*}
where $\boldsymbol{x}_i=(1 ~x_{i1} ~\cdots ~ x_{id} )^T$ and $\boldsymbol{y}=(y_{1} ~y_2~\cdots ~ y_{d} )^T$. The coefficient matrix can be estimated using the method of the ordinary least squares regression method.
\begin{align*}
\boldsymbol{\beta} = (\boldsymbol{X}^T \boldsymbol{X})^{-1}\boldsymbol{X}^T\boldsymbol{y}, 
\end{align*}
where
\begin{align*}
\boldsymbol{\beta} = \begin{pmatrix}
  \beta_{0} \\
  \beta_{1} \\
  \vdots \\
  \beta_{n} \\
\end{pmatrix} \quad \text{and}\quad 
\boldsymbol{X} = \begin{pmatrix}
   1 & x_{11}  & x_{12}  & \cdots & x_{1n} \\
   1 & x_{21}  & x_{22}  & \cdots & x_{2n} \\
    \vdots & \vdots & \vdots & \ddots & \vdots \\
   1 &  x_{d1} & x_{2}  & \cdots & x_{dn}
\end{pmatrix}.
\end{align*}
In this paper, we use the first $12$ variables at the start of the current hour listed in Table \ref{tab:feature_data} to forecast the PC of the current hour (named as lr\_pct in Table \ref{tab:feature_data}). 

\subsubsection{Polynomial Regression}\label{subsec:3.1.2}
In polynomial regression, we assume a $n$-th degree polynomial relationship between dependent and independent variables. This method can be seen as an extension of linear regression, allowing for more complex relationships to be captured. The polynomial regression can be expressed as follows:
\begin{align*}
    \boldsymbol{y} = \beta_0 + \beta_1  \boldsymbol{x} + \beta_2 \boldsymbol{x}^2 + \cdots + \beta_m \boldsymbol{x}^m, 
\end{align*}
where $m$ denotes the degree of polynomial regression. In our analysis, we employ polynomial regression with a degree $m=2$. Specifically, we incorporate the first 12 variables from the beginning of the current hour, as listed in Table \ref{tab:feature_data}. To enhance the dataset, we augment it with quadratic features ($2$nd-degree terms) derived from these variables. Subsequently, we utilize the method of Ordinary Least Squares (OLS) with linear regression to forecast the PC of the current hour (denoted as poly\_pct in Table \ref{tab:feature_data}).

\subsubsection{XGBoost Regression}\label{subsec:3.1.3}
The XGBoost algorithm, introduced by Chen and Guestrin \cite{xgboost}, represents an extension of the gradient-boosting algorithm initially proposed by Friedman \cite{gradboosting}. XGBoost operates as an ensemble method of additive regression trees, which can be expressed as follows:
\begin{align*}
F(x) = \sum_{k=1}^{K} f_k(x).
\end{align*}
Here, $F(x)$ is the target prediction. Each tree, denoted as $f_k(x)$, functions by mapping the input features $x$ to a prediction. These trees are trained sequentially, with each stage involving the addition of a new tree to rectify the errors made by the existing ensemble. Mathematically, this can be expressed as
\begin{align*}
\hat{y}_i^{(t)} = \sum_{k=1}^{t} f_k(x_i), 
\end{align*}
where $\hat{y}_i^{(t)}$ denotes the prediction at the $t$-th iteration. The objective function of XGBoost comprises a sum of a loss function $L$ and a regularization term $\Omega$ that penalizes the complexity of the model:

\begin{align*}
\zeta^{(t)} = \sum_{i=1}^{n} L(y_i, \hat{y}_i^{(t)}) + \sum_{k=1}^{t} \Omega(f_k),
\end{align*}
where $n$ denotes the number of training samples, $y_i$ is the true target value, and $\hat{y}_i^{(t)}$ is the predicted value at  $t$-th iteration. The loss function $L$ measures the difference between the true target value and the predicted value, and $\Omega(f_k)$ is the regularization term for tree $f_k$. The goal of training XGBoost is to find the optimal set of trees $\{f_k\}$ that minimizes the objective function
\begin{align*}
\{f_k\} = \arg\min_{f_k} \zeta^{(t)}.
\end{align*}
This is typically done using a gradient-based optimization algorithm, where the gradient of the objective function with respect to the predictions is used to update the trees in each iteration:
\begin{align*}
f_k = f_k - \eta ~\nabla_{f_k} \zeta^{(t)},
\end{align*}
where $\eta$ is the learning rate, which controls the step size of the optimization algorithm. The optimization continues until a stopping criterion is met, such as reaching a maximum number of iterations or the improvement in the objective function becomes negligible. Chen and Guestrin \cite{xgboost} and  Chen {\em et al.} \cite{xgboost_btc} have used various models along with XGBoost to predict the price trends across various time intervals. \\
Following the methodology of prior regression models, we utilize the initial 12 variables from the onset of the current hour, as detailed in Table \ref{tab:feature_data}, to predict the PC of the current hour (named as xgb\_pct in Table \ref{tab:feature_data}).

\subsubsection{LGBM Regression}
LGBM is a gradient-boosting method, which uses an ensemble learning framework. It was proposed by Ke {\em et. al.} \cite{lgbm}, where they introduced two algorithms, namely, Gradient-based One-Side Sampling (GOSS) and Exclusive Feature Bundling (EFB), to address high dimensionality and large datasets in machine learning. It optimizes memory usage and training time with techniques such as EFB and GOSS. It is a similar algorithm to XGBoost but uses a gradient-based optimization algorithm GOSS to speed up the training process. GOSS samples a small subset of data with large gradients to compute the gradient, while the rest of the data is sampled uniformly. This reduces the computational cost while preserving the accuracy of the gradient estimation.  The input for LBGM regression is the first 12 variables at the start of the current hour listed in Table \ref{tab:feature_data} to forecast the PC of current hour (named as lgbm\_pct in Table \ref{tab:feature_data}).

\noindent
In conclusion, we have introduced six drift terms, namely pct\_change, drift, lr\_pct, poly\_pct, lgbm\_pct, xgb\_pct, which we utilize as replacements for the $\mu$ term in \eqref{eq:3.4} within our Monte Carlo simulations. The only variable remaining unknown in \eqref{eq:3.4} is the $\sigma$ term, which we predict as our volatility term in the subsequent subsection.

\subsection{Diffusion Modeling}
In MJD model, the diffusion term containing $\sigma$ represents the volatility of the asset and $dW(t)$ signifies the increment in a Wiener process that models the random fluctuations in the asset price. Mathematically,
\begin{align}
\sigma_j = \sqrt{\frac{1}{N} \sum_{i=1}^{N} (\delta_{j-i} - \rho_j})^2, \label{eq:3.8}
\end{align}
where $\delta_j$ represents the $i$-th hour return and $\rho_j$ is the average of returns as defined in \eqref{eq:3.5}.\\
In our framework, the forecasting volatility is essential for calculating the diffusion parameter. One prominent method for modeling volatility is the GARCH model. Also, the GJR-GARCH model stands out for its capability to capture asymmetries in volatility, making it especially well-suited for analyzing cryptocurrencies characterized by significant fluctuations. Unlike conventional GARCH models that assume equal effects of positive and negative shocks on volatility, GJR-GARCH introduces the notion of distinct responses to such shocks. This model distinguishes between positive and negative returns, enabling it to provide more precise estimates of future volatility levels in cryptocurrency data. This is particularly crucial given the tendency of cryptocurrencies to experience heightened volatility following negative price movements.

\noindent
The GJR-GARCH model expands upon the traditional GARCH formulation by incorporating an additional variable multiplied by the squared error term in the conditional variance equation. This model operates as a threshold model, where a dummy variable, the indicator function, switches between two states based on the sign of the preceding period's residual. Specifically, the indicator function equals one if the previous period's residual is negative, and zero otherwise. Consequently, the conditional variance follows distinct processes contingent upon the sign of the error terms.

\noindent
Consider $ \varepsilon_t = \sigma_t + z_t $, where $\varepsilon_t$ is a zero-mean white noise process and $z_t$ is standard Gaussian process. The conditional variance of a GJR-GARCH(1,1) process is defined as
\begin{align*}
\sigma^2_t = \omega + \alpha\varepsilon^2_{t-1} + \gamma I_{t-1}\varepsilon^2_{t-1} + \beta  \sigma \omega^2_{t-1},  
\end{align*}
where constraints $\alpha,\beta,\gamma, \omega >0$ with  $ \alpha+ \beta + 0.5\gamma < 1$ and
\begin{align*}
    I_{t-1}=\left\{
    \begin{array}{ll}
        1, &  \text{if $\epsilon_{t-1}<0$;}\\
        0, & \text{otherwise.}
    \end{array}\right.
\end{align*}
In modeling the above-mentioned model, we use log-likelihood that can be used in an optimizer. The optimization process is constrained by the condition $\alpha + \beta + 0.5\gamma < 1$. The initial values for the coefficients are determined by calculating the mean and variance of past returns. Subsequently, the optimal coefficients for the model are identified through maximum likelihood estimation. Mostafa {\em et al.} \cite{GJRgarch} employed similar modeling techniques in their research to forecast Bitcoin price volatility.

\noindent
In addition to GJR-GARCH, we have employed linear regression, polynomial regression, and XGBoost regression to forecast the volatility term, with `vol' designated as the target variable, utilizing the same technique outlined in Subsections \ref{subsec:3.1.1}, \ref{subsec:3.1.2}, and \ref{subsec:3.1.3}. These four forecasted volatility variables are denoted as forc\_vol\_lr, forc\_vol\_poly, forc\_vol\_xgb, and forc\_vol\_gjr. Consequently, we now have five volatility terms available as alternatives to $\sigma$ in \eqref{eq:3.4} for our Monte Carlo simulations.

\subsection{Metrics}
We approach our original price forecasting problem by segmenting it into two distinct components: signal generation and price prediction. In addressing signal generation, we frame it as a classification-based problem, where the efficacy of our approach is gauged by the accuracy of the signals generated. Conversely, for price prediction, we formulate it as a regression problem, employing widely accepted metrics commonly used in regression analysis.
The classification metrics employed include Accuracy, Precision, Recall, and F1-Score, as defined by
\begin{align*}
    \text{Accuracy} &:= \frac{\text{TP} + \text{TN}}{\text{TP} + \text{TN} + \text{FP} + \text{FN}},\\
    &\\
    \text{Precision} &:= \frac{\text{TP}}{\text{TP} + \text{FP}},\\
    &\\
    \text{Recall} &:= \frac{\text{TP}}{\text{TP} + \text{FN}},\\
    &\\
    \text{F1 Score} &:= 2 \times \frac{\text{Precision} \times \text{Recall}}{\text{Precision} + \text{Recall}},
\end{align*}
where TP is True Positive, TN is True Negative, FP is False Positive, and FN is False Negatives. These values can be calculated from the confusion matrix, which compares the true and predicted signals. RMSE and MAPE are given by
\begin{align*}
\text{RMSE} = \sqrt{\frac{1}{n}\sum_{i=1}^{n}(y_i - \hat{y}_i)^2}    
\end{align*}
and
\begin{align*}
\text{MAPE} = \frac{1}{n}\sum_{i=1}^{n}\left|\frac{y_i - \hat{y}_i}{y_i}\right| \times 100\% ,
\end{align*}
where $y_i$ denotes the actual value in future and $\hat{y}_i$ represent the predicted value in future. These metrics have been previously employed in regression and classification analysis in several papers, including Chen \cite{BITCOINML2} and Chen {\em et al.} \cite{xgboost_btc}.

\section{Data Availability and Preprocessing}\label{12:sec3}
We use freely available OHLCV data from the popular cryptocurrency exchange binance.com, focusing on the BTCUSDT pair. Data for other pairs such as ETHUSDT and ADAUSDT can also be obtained for conducting similar simulations. Specifically, we used the Klines spot data for the BTCUSDT pair with the granularity of `1h' from January 2020 to January 2023. The data provided by Binance was organized on a monthly basis. To facilitate our analysis, we initially developed a monthly file processing pipeline that extracts data from each month's file from January 2020 to January 2023, processes the respective file, adds data from the end of the previous month for the computation of variables such as drift, volatility, among others, and merges the data. We also introduce certain new variables such as percent change, drift, delta drift, and volatility of price (refer to \eqref{eq:3.5}, \eqref{eq:3.7}, and \eqref{eq:3.8}). We calculate these variables on the open price of the BTCUSDT pair. The final target variable for the problem is the price of the BTCUSDT pair, but before that, we use various regression algorithms to forecast the percent change in price over the next hour and the volatility of the price as discussed in Section \ref{sec: Methodology}. To remove any forward bias in the prediction of these terms, we shift all columns except the `open' price by 1 hour, so that apart from the open price all columns will be at a delay of 1 and forecast the terms at the start of the hour. We train our models on a monthly basis, refreshing the weights at the start of every new month, and predicting on a rolling basis retraining the model on the data of past month. The rationale behind this is that data older than a few months does not hold relevance as the price of the BTCUSDT pair is very volatile and changes by a large factor and the models we are using assign weights to every factor, and as the value of asset change is large, the weights become obsolete. 

\noindent
For forecasting the percent change terms, models such as linear regression, polynomial regression, XGBoost regression, and LGBM regression are used with the percent change term as the target variable. For predicting the future drift using polynomial regression, linear regression is used by introducing polynomial features in our dataset, that is, a new dataset is introduced by adding degree 2 components of every feature. For forecasting volatility, we use linear regression, polynomial regression, and XGB regression with volatility as our target variable along with volatility forecasting using  GJR-GARCH. The following table highlights the statistical characteristics of both the current data and the forecasted data, along with the delay associated with each term over the entire period.

\begin{longtable}{ |p{0.8cm}|p{3cm}|p{4.7cm}|p{2.5cm}|p{2.5cm}|p{1cm}| }
    % \centering
    \caption{Statistical Features of Data }
    \label{tab:feature_data}\\
     \hline
     \textbf{S.No.} & \textbf{Name of Feature}& \textbf{Description} & \textbf{Mean} & \textbf{Standard Deviation} & \textbf{Delay} \\
     \hline
     $1$ &open  &  \small Price of BTCUSDT pair at the start of the hour. & $29487.54$  & $16703.18$ & $0$\\
     \hline
     $2$ &close  &  \small Price of BTCUSDT pair at the end of the previous hour.   & $29487.53$  & $16703.18$ & $1$ \\
     \hline
     $3$ &high  &  \small Highest price of BTCUSDT pair in the previous hour.   & $29649.26$  & $16802.37$ & $1$\\
     \hline
     $4$ &low  &  \small Lowest price of BTCUSDT pair in the previous hour.   & $29315.07$  & $16598.64$ & $1$\\
     \hline
     $5$ &volume  & \small Volume of BTCUSDT pair traded in the previous hour.    & $4165.86$ & $4762.776$ & $1$\\
     \hline
     $6$ &quote\_volume &   \small Quote Volume of BTCUSDT pair in the previous hour.  & $1.05425 \times 10^8$ & $1.08626 \times 10^8$ & $1$\\
     \hline
     $7$ &count  &  \small Number of trades in BTCUSDT pair in the previous hour 
     & $8902.19$ & $9284.96$ & $1$\\
     \hline
     $8$ &taker\_buy\_volume  & \small  The total volume of buy orders filled by takers in the previous hour.  & $2055.95$ & $2371.02$ & $1$\\
     \hline
     $9$ &pct\_change  &  \small Percent change in open price between current and previous hour.  & $7 \times 10^{-5}$ &  $8.02 \times 10^{-3}$ & $0$\\
     \hline
     $10$ &drift  &  \small Mean of pct\_change of previous 60 hours. & $6.7 \times 10^{-5}$ & $9.73 \times 10^{-4}$ & $0$\\
     \hline
     $11$ &vol  & \small Standard deviation of pct\_change of previous 60 hours.  & $6.928 \times 10^{-3}$  & $4.05 \times 10^{-3}$ & $0$\\
     \hline
     $12$ &del\_drift  & \small Change in drift between previous and current hour. &  $2.67 \times 10^{-8}$	& $1.89 \times 10^{-4}$ & $0$\\
     \hline
     $13$ &lr\_pct  & \small Forecasted value of pct\_change using Linear regression algorithm for the current hour. & $8.8 \times 10^{-5}$ & $3.53 \times 10^{-3}$ & $0$\\
     \hline
     $14$ &poly\_pct  & \small Forecasted value of pct\_change using Polynomial regression algorithm(degree = 2) for the current hour. & $-5.26 \times 10^{-3}$ & $0.12$ & $0$\\
     \hline
     $15$ &xgb\_pct  &  \small Forecasted value of pct\_change using XGB regression algorithm for the current hour.  & $-2.91 \times 10^{-4}$ & $7.91 \times 10^3$ & $0$\\
     \hline
     $16$ &lgbm\_pct  &  \small Forecasted value of pct\_change using LGBM regression algorithm for the current hour.   & $4.5 \times 10^{-5}$ & $4.73 \times 10^3$ & $0$\\
     \hline
     $17$ &forc\_vol\_lr  & \small Forecasted value of vol using Linear regression algorithm for the current hour.  & $6.92 \times 10^{-3}$ & $3.99 \times 10^{-3}$ & $0$\\
     \hline
     $18$ &forc\_vol\_pct  &  Forecasted value of vol using Polynomial regression algorithm(degree = 2) for the current hour. & $7.03 \times 10^{-3}$ & $24.49 \times 10^{-3}$ & $0$\\
     \hline
     $19$ &forc\_vol\_xgb  &  Forecasted value of vol using XGB regression algorithm for the current hour.& $6.56 \times 10^{-3}$ & $2.61 \times 10^{-3}$ & $0$\\
     \hline
     $20$ &forc\_vol\_gjr  &  Forecasted value of vol using GJR-garch algorithm for the current hour. & $7.11 \times 10^{-3}$ & $2.66 \times 10^{-3}$ & $0$\\
     \hline
\end{longtable}

\noindent
Features such as pct\_change, drift, vol, del\_drift are calculated on open price and thus have 0 delay. Since the close price is being used at a delay of 1, it shows the same statistical features as that of the open price with a delay of 1. It is interesting to see that the standard deviation and mean of high price and low price are the highest and lowest, respectively, with the ratio of standard deviation to mean being greater than 0.5 in all cases. Interestingly, the ratio of standard deviation to mean of all volume-related terms is greater than 1. Moreover, the forecasted mean PC of polynomial regression and XGBoost is negative while that of linear regression and LGBM is positive with the mean of PC being positive.

\noindent
The figure presented below illustrates the correlation between the existing and forecasted variables outlined in Table \ref{tab:feature_data}.

\begin{figure}[H]
    \centering
    \includegraphics[scale=0.4]{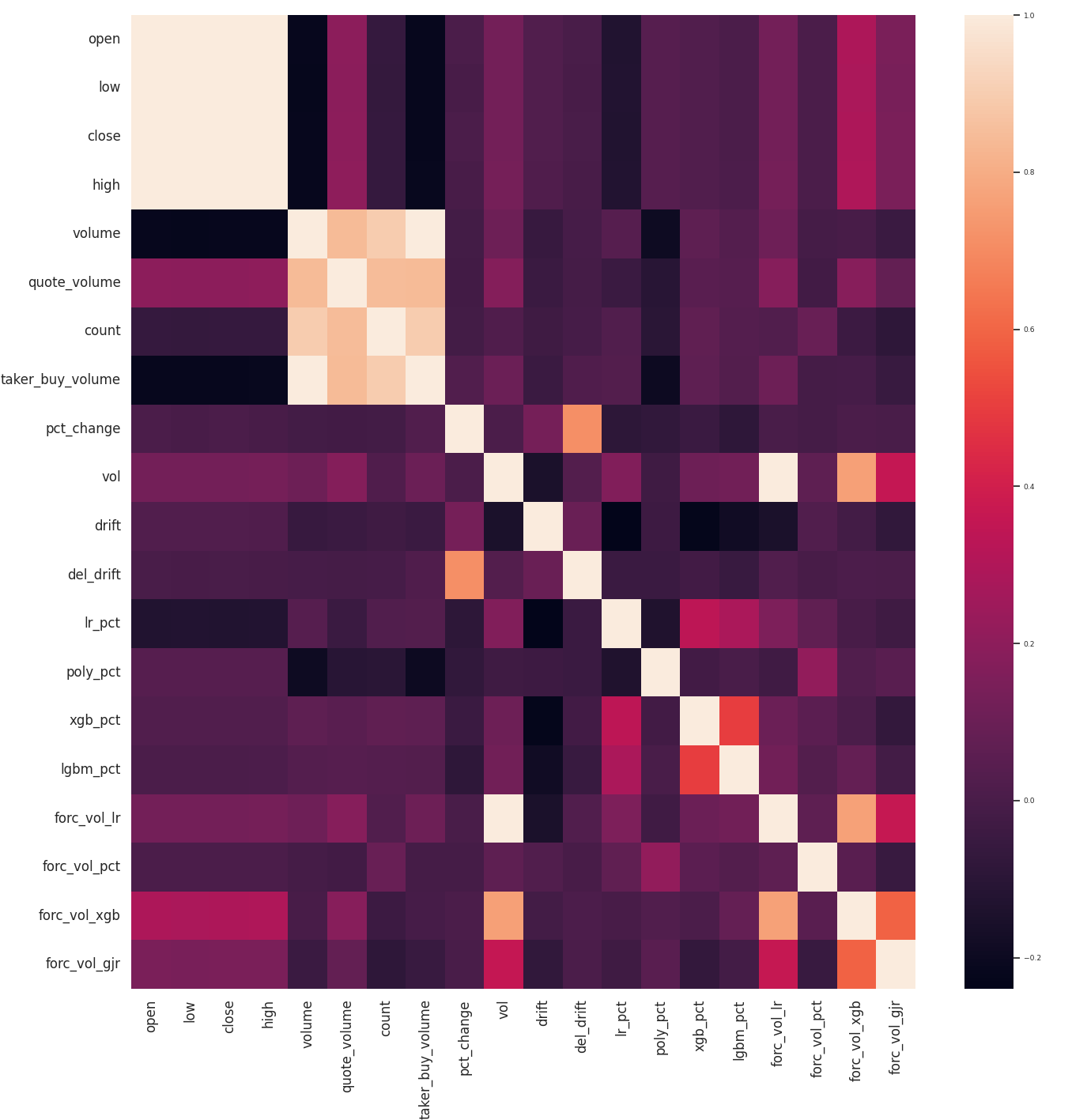}
    \caption{Correlation Heatmap of the Variables}
    \label{fig:corrmap}
\end{figure}

\begin{figure}[H]
    \centering
    \includegraphics[width=\linewidth,height = 0.4\linewidth]{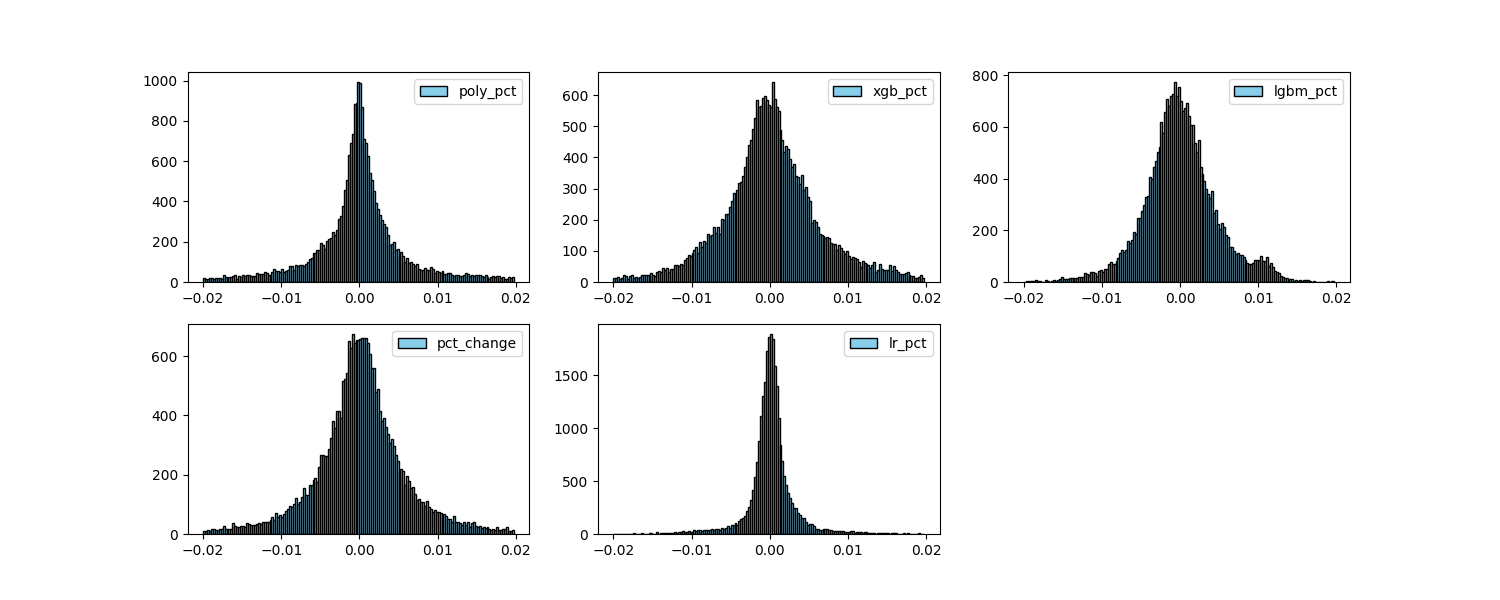}
    \caption{Distribution of Price Change Factors}
    \label{fig:histpct}
\end{figure}

\begin{figure}[H]
    \centering
    \includegraphics[width=\linewidth,height = 0.4\linewidth]{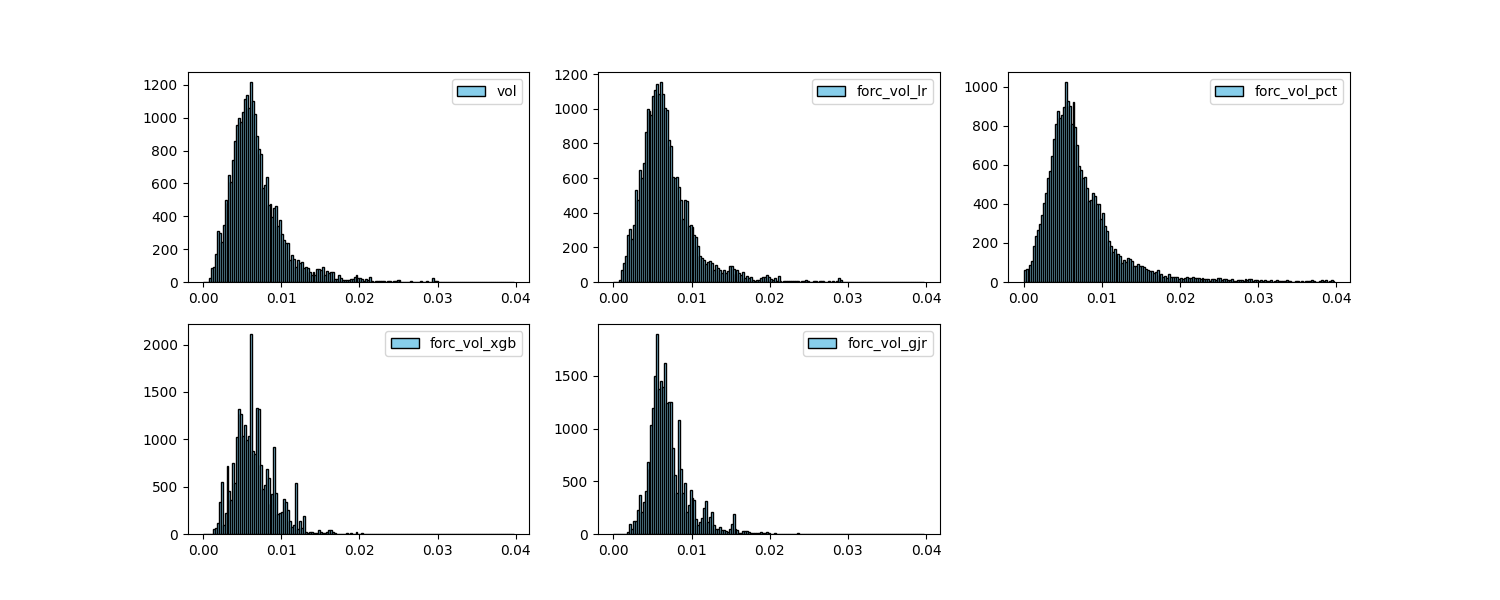}
    \caption{Distribution of Volatility Terms}
    \label{fig:histvol}
\end{figure}

\noindent
The subsequent figures highlight the distribution of percent change, accompanied by the forecasted terms (Figure \ref{fig:histpct}), and the distribution of volatility, along with the forecasted volatility terms (Figure \ref{fig:histvol}).

\noindent
The correlation heatmap (Pearson coefficient in Figure \ref{fig:corrmap}) shows that there is little to no correlation between the percent change and forecasted variables, whereas the distribution of the percent change (Figure \ref{fig:histpct}) can be seen to be very similar to that forecasted by XGBoost and LGBM. In the case of volatility, the forecasted volatility by linear regression is shown to have the highest correlation coefficient with volatility followed by XGBoost, but the distribution of forecasted volatility (Figure \ref{fig:histvol}) by XGBoost is very different than the distribution of actual volatility, which is very similar to the forecasted volatility by linear regression. 

\noindent
Furthermore, it is observed that the forecasted percent change distribution is most accurately modeled by XGBoost and LGBM, as illustrated in Figure \ref{fig:histpct}. Linear regression and polynomial regression, although showing lesser deviation from the actual distribution, still demonstrate relatively good performance compared to other models. In terms of volatility, linear regression and polynomial regression closely approximated the distribution of volatility, while XGBoost and GJR-GARCH exhibited more discrete distributions, as illustrated in Figure \ref{fig:histvol}.

\section{Algorithm and Simulations}\label{12:sec4}
In this section, we discuss an algorithm and simulations to solve MJD model. The figure presented below illustrates the entire flow of operations within the algorithm designed for forecasting the price of the BTCUSDT pair.

    \begin{figure}[H]
        \centering
        \includegraphics[width = 0.85\linewidth]{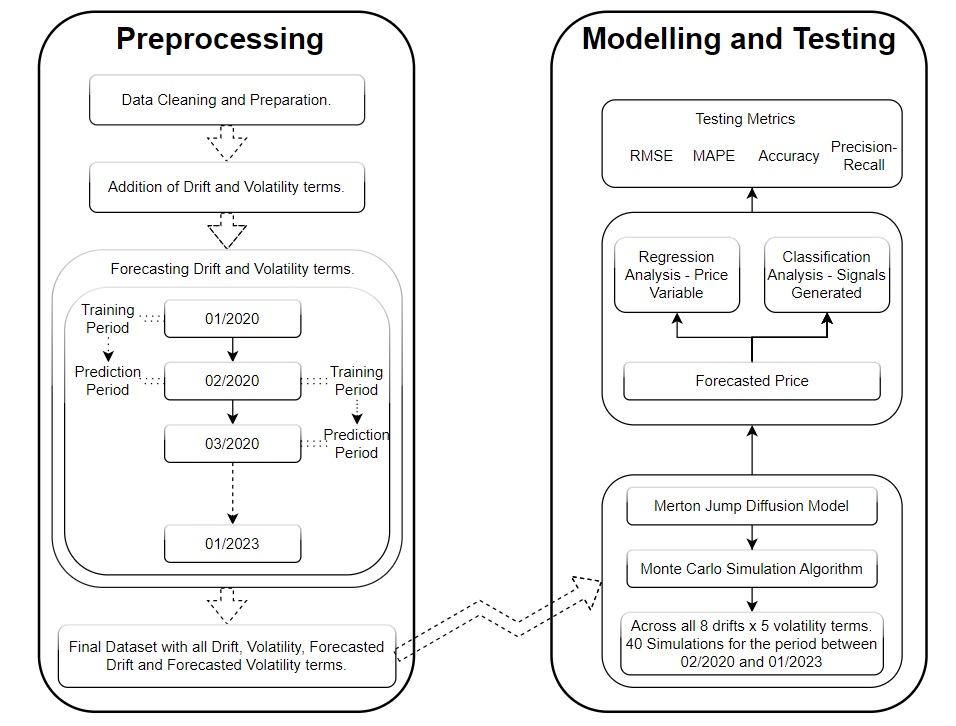}
        \caption{Model for Simulations}
        \label{fig:flowchart}
    \end{figure}

\noindent
For computing the drift term, we utilize \eqref{eq:3.5}, which serves as our generic drift term. Additionally, we implement \eqref{eq:3.5} with a rolling window of $n$ data points (where $n=60$, representing 60 hours in our context). To determine the forecasted drift term, we forecast the drift or find the expected drift of the next hour. It is important to note that we retrain our model weights and biases on a rolling basis with the data from the past month at the start of each month and use this data to forecast the values for the current month. We use the xgboost library for using the eXtreme gradient boosted decision trees-based model, the lgbm library for the LGBM, and sklearn for all the other machine learning models. For the volatility term, we use \eqref{eq:3.8}, which is our generic diffusion term, with a rolling window on $n$ hours ($n=60$ in our case). For forecasting volatility, we again use similar libraries as in drift, along with the computation of expected volatility forecast with the GJR-GARCH model. We compute a total of 6 drift terms and 5 volatility terms that is used in our Monte Carlo simulations as an alternative to the conventional drift and diffusion term, respectively. Along with the existing drift terms, we also introduced 2 new drift terms, the negation of drift and percent change, to test the hypothesis of the reverting nature of cryptocurrency markets. We use Monte Carlo simulation to solve the MJD model given in \eqref{eq:3.1} using \eqref{eq:3.4}. A combination of 40 different drift and volatility pairs (8 drift $\times$ 5 volatility) are tested in place of $\mu$ and $\sigma$, respectively, in \eqref{eq:3.4} for Monte Carlo simulations.

\noindent
We simulate each hour's price forecast in Monte Carlo simulations 10000 times ($N_{sim} = 10000$) to achieve a stable result and take their averages as our final prediction. All simulations are done in Python language using the Jupyter Notebook Framework and by the use of helper libraries numpy, pandas, scipy, statsmodels, sklearn, xgboost, and lgbm. Simulations have been done on the BTCUSDT spot pairs using the spot OHLCV data on an hour-by-hour frequency available on Binance. Depending on the simulation technique, we use a different rolling window for the calculation or forecasting of the drift or diffusion term.  For the jump processes, we assume the jumps arrival distributed by compound Poisson process with the intensity of the jump being equal to twice the drift or the forecasted drift term.

\noindent
A sample simulation of the BTCUSDT spot pair, initiated with a starting price of 16529.59. The drift and diffusion terms are computed using \eqref{eq:3.5} and \eqref{eq:3.8}, respectively, while the jumps are generated through a compound Poisson process. This simulation is depicted in the figure below.

 \begin{figure}[H]
    \centering
    \includegraphics[width = 0.7\linewidth]{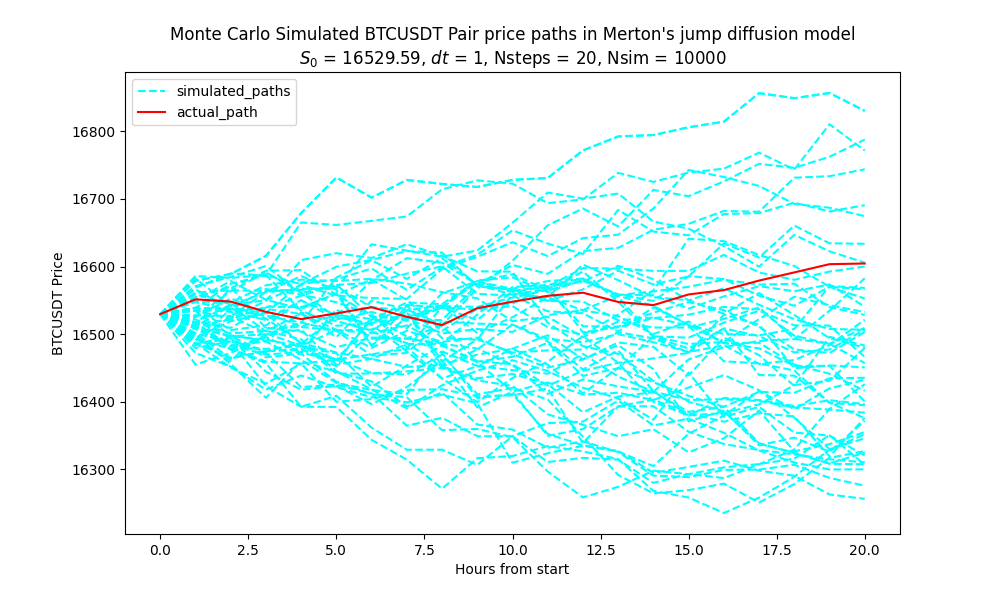}
    \caption{Sample Monte Carlo Simulation}
    \label{fig:montecarlo}
\end{figure}

\noindent
The Monte Carlo simulation provides us with the one-hour ahead price of the BTCUSDT pair, enabling us to generate signals based on the current open price of the pair.  These signals are based on our model comparison to the basis of classification-based metrics (Classification Accuracy, F1 score, Precision, Recall, and Specificity) along with the forecasted price being compared on the basis of regression-based metrics (RMSE and MAPE). Our analysis compares the results both on an hour-by-hour signal generated basis and according to a holding transaction-based principle (comparing related data points only when the signal changes), which is applicable in various real-life applications.

\section{Results and Conclusions}\label{12:sec6}
In this section, we delve into the results obtained and provide pertinent remarks regarding their significance and implications. The table presented below illustrates the results  of various combinations of drift and diffusion terms on regression-based metrics such as RMSE and MAPE based on hour-by-hour as well as transactional levels (denoted by T).

{\small{
\begin{longtable}{|c|c|c|c|c|c|c|}
    % \centering
    \caption{Results retalted to Regression Analysis}
    \label{tab:reg_table}\\
     \hline
     S.No. & Drift Term & Diffusion Term & RMSE & MAPE & RMSE-T & MAPE-T\\
     \hline
     1 & drift\_negated & forc\_vol\_gjr & 264.38 & 0.503 & 539.95 & 0.972
     \\
     2 & drift & forc\_vol\_xgb & 265.60 & 0.507 & 527.39 & 0.956
     \\
     3 & lgbm\_pct & forc\_vol\_lr & 324.99 & 0.675 & 434.07 & 0.811
     \\
     4 & xgb\_pct & forc\_vol\_gjr & 381.05 & 0.793 & 451.33  & 0.825
     \\
     5 & pct\_change\_negated& forc\_vol\_lr & 400.07 & 0.770 & 449.61  & 0.813
     \\
     6 & pct\_change & forc\_vol\_lr & 406.42 & 0.803 & 456.37 & 0.920 \\
     \hline
\end{longtable}
}}
\noindent
Across all 40 different combination pairs of drift and volatility terms, the above-listed terms were either the best in the specified metric or showed interesting conclusions. From the above table, we can see that the pairs (drift\_negated, GJR-garch) and (drift, XGBoost), representing combinations of drift and diffusion terms, have significantly batter values of RMSE and MAPE compared to the other combinations and their overall metric values are quite comparable. The pair (LGBM, linear regression) exhibits the most favorable RMSE-T and MAPE-T values, which are comparable to those achieved by the pairs (XGBoost, GJR-garch) and (pct\_change\_negated, linear regression). The results of  pct\_change\_negated and pct\_change as drift terms with linear regression as diffusion term are very similar for regression-based analysis. Among all the pairs evaluated, polynomial regression (degree 2) exhibited the poorest performance across all volatility terms. It is interesting to observe that the results concerning different volatility terms with fixed drift term show similar distributions.  The combination of linear regression and GJR-GARCH often yielded the most favorable outcomes, while polynomial regression tends to produce the least favorable results in regression analysis. 
%It doesn't make much sense to compare the RMSE result of our model with different papers as the period of testing used in the available papers is very different from the period used in our paper.

\noindent
The figure presented below illustrates the actual price movement of the BTCUSDT pair from January 27, 2023, to January 31, 2023, alongside the forecasted prices derived from drift terms, keeping the diffusion term as vol.
 \begin{figure}[H]
    \centering
    \includegraphics[width = 1\linewidth]{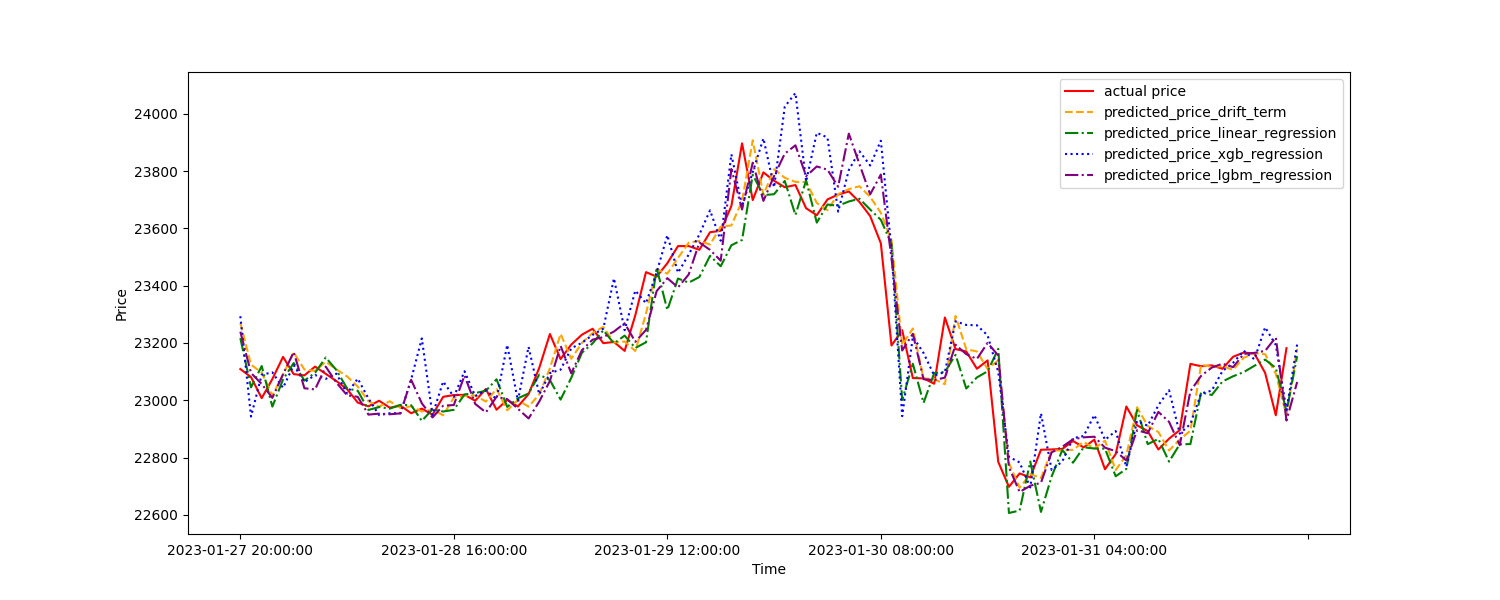}
    \caption{Predicted Prices with Various Drift Terms and Volatility}
    \label{fig:price_plot}
\end{figure}

\noindent
Although the figure represents only a small portion of the entire dataset, certain conclusions can be drawn. For instance, it is evident that the predicted price forecast generated by XGBoost exhibits the highest deviation from the actual price, while various other drift terms show minimal deviation. These conclusions are concluded by both Table \ref{tab:reg_table} and Figure \ref{fig:price_plot}.

\noindent
The following table shows the results of various combinations of drift and diffusion terms on classification based metrics like Accuracy, Precision, Recall, and F1-Score.

{\footnotesize{
\begin{longtable}{|c|c|c|c|c|c|c|c|}
    % \centering
    \caption{Results related to Classification Analysis}
    \label{tab:class_analysis}\\
     \hline
     S.No. & Drift Term & Diffusion Term & Accuracy & Accuracy - T & Precision-T & Recall-T & F1-score-T \\
     \hline
     1 & pct\_change\_negated & forc\_vol\_lr & 0.527 & 0.649 & 0.663& 0.645 & 0.654
     \\
     2 & drift\_negated & forc\_vol\_gjr  & 0.508 & 0.613	& 0.614	&0.612	&0.613
     \\
     3 & lr\_pct & forc\_vol\_gjr & 0.504 &0.530 &0.547 &0.52 &0.538
     \\
     4 & xgb\_pct & forc\_vol\_xgb & 0.508 & 0.530 & 0.539 & 0.529 & 0.534
     \\
     5 & lgbm\_pct & forc\_vol\_lr & 0.504 & 0.512 & 0.529 & 0.511 & 0.520
     \\
     6 & poly\_pct & forc\_vol\_poly & 0.497	& 0.510	 & 0.524 & 0.507 & 0.515
     \\
     7 & drift & forc\_vol\_poly & 0.494 & 	0.393 &	0.404 &	0.397 &	0.401
     \\
     8 & pct\_change & forc\_vol\_poly & 0.473 & 0.351 & 0.363 &0.354 & 0.358\\
     \hline
\end{longtable}}}

\noindent
The combinations of drift and diffusion pairs presented in the table above offer valuable insights into classification analysis, particularly concerning signal generation derived from predicted price movements. The table presents the optimal pair for each drift term in classification analysis across all diffusion terms. Interestingly, the pair (pct\_change\_negated, linear regression) has one of the best MAPE and RMSE on a transactional basis.  Furthermore, it demonstrated the highest transactional accuracy along with signal accuracy, indicating a substantial enhancement in capturing data trends compared to all other pairs. Moreover,  it suggests that the models maintained stability over the observed period as the transactional accuracy, precision, and recall remained consistent across all the models.  The expected inverse relationship between the results of pct\_change and pct\_change\_negated aligns well with our hypothesis, as evidenced by the data presented in Table \ref{tab:class_analysis}. Among all ML models, polynomial regression performed the worst in classification as well as regression analysis in the modeling of the drift term. The results from both gradient boosted tree-based methods were comparable, with XGBoost exhibiting a slight advantage over LGBM in terms of transactional accuracy. The GJR-GARCH model consistently emerges as one of the top performers in volatility modeling.

\noindent
The pair of negated drift along with GJR-GARCH emerged as particularly noteworthy in both the regression and classification analyses, displaying remarkably low RMSE and MAPE scores, along with consistently high accuracy rates compared to all other pairs. In this study, we investigated MJD model coupled with various machine learning and statistical techniques to predict Bitcoin price. 
Most of the previous work in the field has concentrated on the use of machine learning (McNally {\em et al.} \cite{BITCOINML1}), machine learning with added features from various sources such as data from other asset classes (Chen \cite{BITCOINML2}) or the inclusion of high-dimensional features across high frequency and daily time frames (Chen {\em et al.} \cite{xgboost_btc}). The results suggest that statistical methods outperform classification-oriented approaches, whereas machine learning models excel in regression-based tasks. We envision that our approach, which involves utilizing the MJD Model alongside forecasting drift and diffusion parameters using machine learning models for price prediction, can be extended to various domains requiring the simulation of differential equations for solutions and other areas related to cryptocurrency or the stock market with a similar modeling problem. 

\noindent
Our study has various limitations regarding data sources and analyses, suggesting potential extensions for future research. We rely on a single source for price volume data of fixed granularity for the prediction of price. For a comprehensive analysis of Bitcoin price prediction, it would be advantageous to collect price data spanning various granularities and incorporate features with greater dimensionality. We aim to consider other features and focus on incorporating more useful elements such as data from various other cryptocurrencies, other asset classes, and sentiment analysis of social networks for forecasting drift and diffusion terms in further studies.

\section*{Acknowledgements}
The first two authors are supported by Institute Fellowship from IIT (BHU) Varanasi, India. The third author is partially supported by SERB Start-up Research Grant (File No. SRG/2023/000236) and MATRICS Research Grant (File No. MTR/2023/000200), India.

\setstretch{1.18}
%\singlespacing
\small
%\footnotesize
\bibliographystyle{PV}

\bibliography{PA2PSDBib}

\end{document}